\documentclass[aps,amssymb,twocolumn]{revtex4}
\usepackage{amssymb}
\usepackage{graphicx}
\usepackage{color}
\usepackage{textcomp}
\usepackage{ulem}
\begin{document}

\title{Inverse magnetic catalysis effect and current quark mass effect on mass spectra and Mott transitions of pions under external magnetic field}
\author{Luyang Li$^1$}
\author{Shijun Mao$^2$}%
 \email{maoshijun@mail.xjtu.edu.cn}
\affiliation{$^1$ Xi'an University of Posts and Telecommunications, Xi'an, Shaanxi 710121, China\\
$^2$ Institute of Theoretical Physics, School of Physics, Xi'an Jiaotong University, Xi'an, Shaanxi 710049, China}

\begin{abstract}
Mass spectra and Mott transition of pions $(\pi^0,\ \pi^\pm)$ at finite temperature and magnetic field are investigated in a two-flavor NJL model, and we focus on the inverse magnetic catalysis (IMC) effect and current quark mass (CQM) effect. Due to the dimension reduction of the constituent quarks, the pion masses jump at their Mott transitions, which is independent of the IMC effect and CQM effect. We consider the IMC effect by using a magnetic dependent coupling constant, which is a monotonic decreasing function of magnetic field. With IMC effect, the Mott transition temperature of $\pi^0$ meson $T_m^0$ is a monotonic decreasing function of magnetic field. For charged pions $\pi^{\pm}$, the Mott transition temperature $T_m^+$ fast increases in weak magnetic field region and then decreases with magnetic field, which are accompanied with some oscillations. Comparing with the case without IMC effect, $T_m^0$ and $T_m^+$ are lower when including IMC effect. CQM effect are considered by varying parameter $m_0$ in non-chiral limit. For $\pi^0$ meson, $T_m^0$ is not a monotonic function of magnetic field with low $m_0$, but it is a monotonic decreasing function with larger $m_0$. In the weak magnetic field region, $T_m^0$ is higher for larger $m_0$, but in the strong magnetic field region, it is lower for larger $m_0$. For $\pi^+$ meson, $T^+_m$ is only quantitatively modifies by current quark mass effect, and it becomes higher with larger $m_0$.
\end{abstract}

\date{\today}
\maketitle

\section{Introduction}
The study of hadron properties in QCD medium is important for our understanding of strong interaction matter, due to its close relation to QCD phase structure and relativistic heavy ion collision. For instance, the chiral symmetry breaking leads to the rich meson spectra, and the mass shift of hadrons will enhance or reduce their thermal production in relativistic heavy ion collisions~\cite{k1,k2,k3}.
	
It is widely believed that the strongest magnetic field in nature may be generated in the initial stage of relativistic heavy ion collisions. The initial magnitude of the field can reach $eB\sim (1-100)m_\pi^2$ in collisions at the Relativistic Heavy Ion Collider and the Large Hadron Collider~\cite{b0,b1,b2,b3,b4}, where $e$ is the electron charge and $m_\pi$ the pion mass in vacuum. In recent years, the study of magnetic field effects on the hadrons attract much attention. As the Goldstone bosons of the chiral (isospin) symmetry breaking, the properties of neutral (charged) pions at finite magnetic field, temperature and density are widely investigated~\cite{c1,c3,hadron1,hadron2,qm1,qm2,sigma1,sigma2,sigma3,sigma4,l1,l2,l3,l4,lqcd5,lqcd6,ding2008.00493,ding2022,njl2,meson,mfir,ritus5,ritus6,mao1,mao11,mao2,wang,coppola,phi,liuhao3,he,Restrepo,maocharge,maopion,yulang2010.05716,q1,q2,q3,q4,huangamm1,huangamm2,q5,q6,q7,q8,q9,q10,su3meson4,Scoccola1,Scoccola2,meijie3}. 

The LQCD simulations performed with physical pion mass observe the inverse magnetic catalysis (IMC) phenomenon~\cite{lattice1,lattice11,lattice2,lattice3,lattice4,lattice5,lattice6,lattice7}. Namely,
the pseudo-critical temperature $T_{pc}$ of chiral symmetry restoration and the quark mass near $T_{pc}$ drop down with increasing magnetic field. On analytical side, many scenarios are proposed to understand this IMC phenomenon, but the physical mechanism is not clear~\cite{fukushima,mao,hadron2,bf1,bf12,bf13,bf2,bf3,bf4,bf5,bf51,bf52,bf8,bf9,bf11,db1,db2,db3,db5,db6,pnjl1,pnjl2,pnjl3,pnjl4,pqm,ferr1,ferr2,huangamm2,meijie,imcreview,ammmao}.
Besides, how does the IMC effect influence the pion properties under external magnetic field? Previous studies focus on the case with vanishing temperature and density, and report that IMC effect leads to a lower mass for pions under external magnetic field~\cite{meson,su3meson4,mao11}.

The current quark mass determines the explicit breaking of chiral symmetry. In LQCD simulations, it is considered through different pion mass~\cite{l4,lattice5,lattice3}. At vanishing temperature and finite magnetic field, the normalized neutral pion mass $m_{\pi^0}(eB)/m_{\pi^0}(eB=0)$ increases with current quark mass, but the normalized charged pion mass $m_{\pi^\pm}(eB)/m_{\pi^\pm}(eB=0)$ decreases with current quark mass~\cite{l4}. In effective models, the current quark mass (CQM) effect on pion properties has not yet been studied.

In this paper, we will investigate the IMC effect and CQM effect on mass spectra and Mott transition of pions at finite temperature and magnetic field. Here we make use of a Pauli-Villars regularized Nambu--Jona-Lasinio (NJL) model, which describes remarkably well the static properties of light mesons~\cite{njl1,njl2,njl3,njl4,njl5,zhuang}. In our calculations, the IMC effect is introduced by a magnetic field dependent coupling, and the CQM effect is considered by tuning the quark mass parameter.

The rest of paper is organized as follows. Sec.\ref{sec:f} introduces our theoretical framework for meson spectra and the Mott transition in a Pauli-Villars regularized NJL model. The numerical results and discussions are presented in Sec.\ref{sec:r}, which focus on inverse magnetic catalysis effect in Sec.\ref{sec:r}A and current quark mass effect in Sec.\ref{sec:r}B. Finally, we give the summary in Sec.\ref{sec:s}.

\section{Framework}
\label{sec:f}
The two-flavor NJL model is defined through the Lagrangian density in terms of quark fields $\psi$~\cite{njl1,njl2,njl3,njl4,njl5,zhuang}
\begin{equation}
\label{njl}
{\cal L} = \bar{\psi}\left(i\gamma_\nu D^\nu-m_0\right)\psi+G \left[\left(\bar\psi\psi\right)^2+\left(\bar\psi i\gamma_5{\vec \tau}\psi\right)^2\right].
\end{equation}
Here the covariant derivative $D_\nu=\partial_\nu+iQ A_\nu$ couples quarks with electric charge $Q=diag (Q_u,Q_d)=diag (2e/3,-e/3)$ to the external magnetic field ${\bf B}=(0, 0, B)$ in $z$-direction through the potential $A_\nu=(0,0,Bx_1,0)$. $m_0$ is the current quark mass, which determines the explicit breaking of chiral symmetry. $G$ is the coupling constant in scalar and pseudo-scalar channels, which determines the spontaneously breaking of chiral symmetry and isospin symmetry.

In NJL model, mesons are constructed through quark bubble summations in the frame of random phase approximation~\cite{njl2,njl3,njl4,njl5,zhuang},
\begin{equation}
{\cal D}_M(x,z)  = 2G \delta(x-z)+\int d^4y\ 2G \Pi_M(x,y) {\cal D}_M(y,z),
\label{dsequ}
\end{equation}
where ${\cal D}_M(x,y)$ represents the meson propagator from $x$ to $y$ in coordinate space, and the corresponding meson polarization function is the quark bubble,
\begin{equation}
\label{bubble}
\Pi_M(x,y) = i{\text {Tr}}\left[\Gamma_M^{\ast} S(x,y) \Gamma_M  S(y,x)\right]
\end{equation}
with the meson vertex
\begin{equation}
\label{vertex} \Gamma_M = \left\{\begin{array}{ll}
1 & M=\sigma\\
i\tau_+\gamma_5 & M=\pi_+ \\
i\tau_-\gamma_5 & M=\pi_- \\
i\tau_3\gamma_5 & M=\pi_0\ ,
\end{array}\right.
\Gamma_M^* = \left\{\begin{array}{ll}
1 & M=\sigma\\
i\tau_-\gamma_5 & M=\pi_+ \\
i\tau_+\gamma_5 & M=\pi_- \\
i\tau_3\gamma_5 & M=\pi_0\ ,
\end{array}\right. \nonumber
\end{equation}
the quark propagator matrix in flavor space $S=diag(S_u,\ S_d)$, and the trace in spin, color and flavor spaces.

According to the Goldstone's theorem, the pseudo-Goldstone mode of chiral (isospin) symmetry breaking under external magnetic field is the neutral pion $\pi^0$ (charged pions $\pi^\pm$)~\cite{gold1,gold2}. The charged pions are no longer the pseudo-Goldstone modes since their direct interaction with magnetic field.

\subsubsection{neutral pion $\pi^0$}
The neutral pion $\pi^0$ is affected by external magnetic field only through the pair of charged constituent quarks, and its propagator in momentum space can be derived as
\begin{equation}
\label{npole}
{\cal D}_{\pi^0}(k)=\frac{2G}{1-2G\Pi_{\pi^0}(k)},
\end{equation}
with the polarization function $\Pi_{\pi^0}(k)$ and conserved momentum $k=(k_0, {\bf k})$ of $\pi^0$ meson under external magnetic field.

The meson pole mass $m_{\pi^0}$ is defined as the pole of the propagator at zero momentum ${\bf k}={\bf 0}$,
\begin{equation}
\label{mmass}
1-2G\Pi_{\pi^0}(\omega^2=m^2_{\pi^0}, {\bf k}^2=0)=0.
\end{equation}
At nonzero magnetic field, the three-dimensional quark momentum integration in the polarization function $\Pi_{\pi^0}$ becomes a one-dimensional momentum integration and a summation over the discrete Landau levels. The polarization function can be simplified
\begin{eqnarray}
\label{pi}
\Pi_{\pi^0}(\omega^2,0) = J_1(m_q)+\omega^2 J_2(\omega^2)
\end{eqnarray}
and
\begin{eqnarray}
J_1(m_q) &=& 3\sum_{f,n}\alpha_n \frac{|Q_f B|}{2\pi} \int \frac{d p_3}{2\pi} \frac{1-2F(E_f)}{ E_f},\nonumber \\
J_2(\omega^2) &=& 3\sum_{f,n}\alpha_n \frac{|Q_f B|}{2\pi} \int \frac{d p_3}{2\pi}{{1-2F(E_f)}\over  E_f (4 E_f^2-w^2)},\nonumber
\end{eqnarray}
with the summation over all flavors and Landau energy levels, spin factor $\alpha_n=2-\delta_{n0}$, quark energy $E_f=\sqrt{p^2_3+2 n |Q_f B|+m_q^2}$, and Fermi-Dirac distribution function $F(x)=\left( e^{x/T}+1 \right)^{-1}$.

The (dynamical) quark mass $m_q$ is determined by the gap equation,
\begin{eqnarray}
1-2GJ_1(m_q)&=&\frac{m_0}{m_q}.
\label{gap}
\end{eqnarray}

During the chiral symmetry restoration, the quark mass decreases, and the $\pi^0$ mass increases, as guaranteed by the Goldstone's theorem~\cite{gold1,gold2}. When the $\pi^0$ mass is beyond the threshold
\begin{equation}
m_{\pi^0} = 2m_q,
\end{equation}
the decay channel $\pi^0 \rightarrow q {\bar q}$ opens, which defines the ${\pi^0}$ Mott transition~\cite{mott1,zhuang,mott2,mott3}.

From the explicit expression of $\Pi_{\pi^0}$ in Eq(\ref{pi}), the factor $1/(4E_f^2-\omega^2)$ in the integrated function of $J_2(\omega^2)$ becomes $(1/4)/(p_3^2+2n|Q_f B|)$ at $\omega= 2 m_q$. When we do the integration over $p_3$, the $p_3^2$ in the denominator leads to the infrared divergence at the lowest Landau level $n=0$. Therefore, $m_{\pi^0}= 2 m_q$ is not a solution of the pole equation, and there must be a mass jump for $\pi^0$ meson at the Mott transition. This mass jump is a direct result of the quark dimension reduction~\cite{mao2,maocharge,q2,q3,yulang2010.05716}, and independent of the parameters and regularization schemes of NJL model. When the magnetic field disappears, there is no more quark dimension reduction, the integration
$\int d^3{\bf p}/(4E_f^2-\omega^2)\sim \int dp$ becomes finite at $\omega = 2m_q$, and there is no more such a mass jump. It should be mentioned that in chiral limit, no such mass jump happens for ${\pi^0}$ meson even under external magnetic field, which has been analytically proved in our previous work~\cite{mao1}.

\subsubsection{charged pions $\pi^\pm$}
When constructing charged mesons through quark bubble summations, we should take into account of the interaction between charged mesons and magnetic fields. The charged pions $\pi^\pm$ with zero spin are Hermite conjugation to each other, they have the same mass at finite temperature and magnetic field.

The $\pi^+$ meson propagator $D_{\pi^+}$ can be expressed in terms of the polarization function $\Pi_{\pi^+}$~\cite{maocharge,mao11,maopion},
\begin{eqnarray}
\label{eq4}
D_{\pi^+}({\bar k})=\frac{2G}{1-2G\Pi_{\pi^+}({\bar k})},
\end{eqnarray}
where ${\bar k} =(k_0,0,-\sqrt{(2l+1)eB},k_3)$ is the conserved Ritus momentum of $\pi^+$ meson under magnetic fields.

The meson pole mass $m_{\pi^+}$ is defined through the pole of the propagator at zero momentum $(l=0,\ k_3=0)$,
\begin{eqnarray}
1-2G\Pi_{\pi^+}(k_0=m_{\pi^+})=0,
\label{pip}
\end{eqnarray}
and
\begin{eqnarray}
\Pi_{\pi^+}(k_0) &=& J_1(m_q)+J_3(k_0),\label{eq7}\\
J_3(k_0) &=& \sum_{n,n'} \int \frac{d p_3}{2\pi}\frac{j_{n,n'}(k_0)}{4E_n E_{n'}} \times \label{eq8}\\
&& \left[\frac{F(-E_{n'})- F(E_n)}{k_0+E_{n'}+E_n}+\frac{F(E_{n'})- F(-E_n)}{k_0-E_{n'}-E_n}\right],\nonumber\\
j_{n,n'}(k_0)& = &\left[{(k_0)^2/2}-n'|Q_u B|-n|Q_d B|\right]j^+_{n,n'} \nonumber\\
&&-2 \sqrt{n'|Q_u B|n|Q_d B|}\ j^-_{n,n'},\label{eq9}
\end{eqnarray}
with $u$-quark energy $E_{n'}=\sqrt{p^2_3+2 n' |Q_u B|+m_q^2}$, $d$-quark energy $E_n=\sqrt{p^2_3+2 n |Q_d B|+m_q^2}$, and summations over Landau levels of $u$ and $d$ quarks in $J_3(k_0)$.

The quark dimension reduction also leads to infrared ($p_3\to 0$) singularity of the quark bubble $\Pi_{\pi^+}(k_0)$ at some Landau level and some temperature~\cite{mao1,mao2}, and thus there is no solution of the corresponding pole equation for the $\pi^+$ meson mass $m_{\pi^+}$ in this case. Because the spins of $u$ and ${\bar d}$ quarks at the lowest-Landau-level are always aligned parallel to the external magnetic field, and $\pi^+$ meson has spin zero. The two constituents at the lowest Landau level ($n=n'=0$) do not satisfy the pole equation (\ref{pip}). Namely, they can not form a charged $\pi^+$ meson. Considering $|Q_d| < |Q_u|$, the threshold for the singularity of $\Pi_{\pi^+}$ is located at Landau levels $n'=0$ and $n=1$,
\begin{eqnarray}
m_{\pi^+} = m_q+\sqrt{2{|Q_dB|}+m_q^2}.
\end{eqnarray}
This defines the Mott transition of $\pi^+$ meson, and a mass jump will happen. There exist other mass jumps located at $n'\geq 1, n\geq 0$, see examples in Fig.\ref{masspicharget} and Fig.\ref{fmpichargem0}. All these mass jumps are caused by the quark dimension reduction under external magnetic field, and do not depend on the parameters and regularization schemes of NJL model.

\section{results and discussions}
\label{sec:r}
Because of the four-fermion interaction, the NJL model is not a renormalizable theory and needs regularization. In this work, we make use of the gauge invariant Pauli-Villars regularization scheme~\cite{njl1,njl2,njl3,njl4,njl5,mao,maocharge,mao2}, where the quark momentum runs formally from zero to infinity. The three parameters in the Pauli-Villars regularized NJL model, namely the current quark mass $m_0=5$ MeV, the coupling constant $G=3.44$ GeV$^{-2}$ and the Pauli-Villars mass parameter $\Lambda=1127$ MeV are fixed by fitting the chiral condensate $\langle\bar\psi\psi\rangle=-(250\ \text{MeV})^3$, pion mass $m_\pi=134$ MeV and pion decay constant $f_\pi=93$ MeV in vacuum with $T=\mu=0$ and $eB=0$. In our current calculations, we consider the situation with finite temperature and magnetic field and vanishing quark chemical potential $\mu=0$.

\subsection{Inverse magnetic catalysis effect}
From LQCD simulations, the inverse magnetic catalysis phenomenon can be characterized either by the chiral condensates or the pseudo-critical temperature of chiral symmetry restoration~\cite{lattice1,lattice11,lattice2,lattice3,lattice4,lattice5,lattice6,lattice7}. Therefore, to include the inverse magnetic catalysis effect in the NJL model, one approach is to fit the LQCD results of chiral condensates~\cite{geb1,meson,geb3,geb4}, and another approach is to fit the LQCD result of pseudo-critical temperature~\cite{bf8,bf9,geb1,su3meson4,mao11}.

\begin{figure}[hbt]
\centering
\includegraphics[width=7cm]{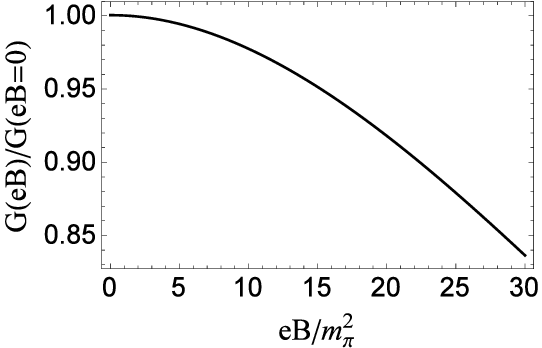}
\caption{Magnetic field dependent coupling $G(eB)$ fitted from LQCD reported decreasing pseudo-critical temperature of chiral symmetry restoration $T_{pc}(eB)/T_{pc}(eB=0)$.} \label{geb}
\end{figure}

In our calculations, following Refs~\cite{bf8,bf9,geb1,su3meson4,mao11}, we use a two-flavor NJL model with a magnetic field dependent coupling $G(eB)$, which is determined by fitting the LQCD reported decreasing pseudo-critical temperature of chiral symmetry restoration $T_{pc}(eB)/T_{pc}(eB=0)$~\cite{lattice1}. As plotted in Fig.\ref{geb}, the magnetic field dependent coupling $G(eB)/G(eB=0)$ is a monotonic decreasing function of magnetic field, and it reduces $16\%$ at $eB/m^2_{\pi} = 30$. In this paper, we fix $m_{\pi}=134$ MeV as the scale of magnetic field. As we have checked, with our fitted coupling constant $G(eB)$, the magnetic catalysis phenomena of chiral condensates at low temperature and the inverse magnetic catalysis phenomena at high temperature can be reproduced.

\subsubsection{neutral pion $\pi^0$}
\begin{figure}[hbt]
\centering
\includegraphics[width=7cm]{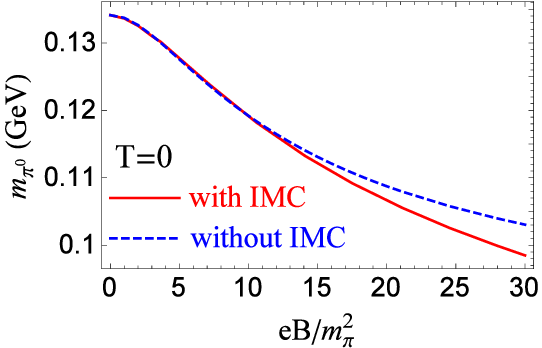}
\caption{$\pi^0$ meson mass $m_{\pi^0}$ as a function of magnetic field at vanishing temperature, with (red solid line) and without (blue dashed line) inverse magnetic catalysis effect.} \label{masspi0}
\end{figure}

With our fitted $G(eB)$ in Fig.\ref{geb}, we solve the gap equation (\ref{gap}) and pole equation (\ref{mmass}) to obtain the $\pi^0$ meson mass at finite temperature and magnetic field with IMC effect. The results with and without IMC effect are plotted in red and blue, respectively.

Figure \ref{masspi0} depicts the $\pi^0$ meson mass $m_{\pi^0}$ as a function of magnetic field at vanishing temperature with (red solid line) and without (blue dashed line) inverse magnetic catalysis effect. Because in both cases the magnetic field enhances the breaking of chiral symmetry in vacuum. As the pseudo-Goldstone boson, the $\pi^0$ meson masses are decreasing functions of magnetic field with and without IMC effect. We observe a lower value for $m_{\pi^0}$, when including the IMC effect. Similar conclusion is obtained in Refs~\cite{meson,su3meson4}, where inverse magnetic catalysis effect is introduced into the two-flavor and three-flavor NJL models.

\begin{figure}[hbt]
\centering
\includegraphics[width=7cm]{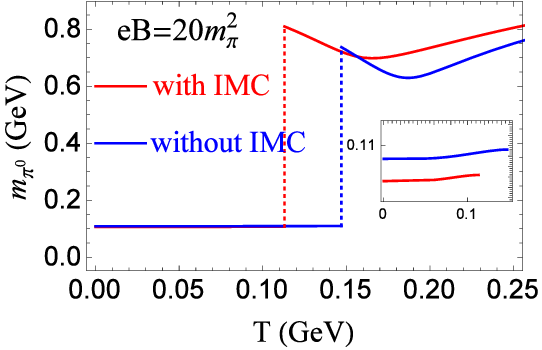}
\caption{$\pi^0$ meson mass $m_{\pi^0}$ as a function of temperature under fixed magnetic field $eB=20 m_{\pi}^2$, with (red lines) and without (blue lines) inverse magnetic catalysis effect.} \label{masspi0t}
\end{figure}

In Fig.\ref{masspi0t}, we plot the $\pi^0$ meson mass $m_{\pi^0}$ at finite temperature and fixed magnetic field $eB=20 m_{\pi}^2$. With (red lines) and without (blue lines) inverse magnetic catalysis effect, the $\pi^0$ mass spectra have the similar structure. $m_{\pi^0}$ increases at low temperature region and jumps at the Mott transition $T=T^0_m$. After that, it firstly decreases and then increases with temperature. The inverse magnetic catalysis effect leads to some quantitative modifications. For instance, Mott transition temperature $T^0_m$ is shifted to a lower value, and $m_{\pi^0}$ becomes lower (higher) at $T<T^0_m$ ($T>T^0_m$).

\begin{figure}[hbt]
\centering
\includegraphics[width=7cm]{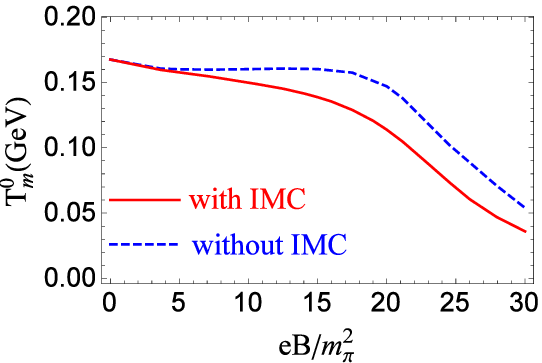}
\caption{Mott transition temperature of $\pi^0$ meson $T^0_m$ as a function of magnetic field, with (red solid line) and without (blue dashed line) inverse magnetic catalysis effect.} \label{tmottpi0}
\end{figure}

Figure \ref{tmottpi0} shows the Mott transition temperature of $\pi^0$ meson $T^0_m$ as a function of magnetic field, with (red solid line) and without (blue dashed line) inverse magnetic catalysis effect. Without IMC effect, the $T^0_m$ decreases with the magnetic field in weak magnetic field region $(eB\leq 5 m^2_{\pi})$, becomes flat in medium magnetic field region $(5 m^2_{\pi} \leq eB \leq 16 m^2_{\pi})$ and decreases in strong magnetic field region $(eB\geq 16 m^2_{\pi})$. With IMC effect, the flat structure of Mott transition temperature disappears and $T^0_m$ monotonically decreases with magnetic field. Furthermore, at fixed magnetic field, $T^0_m$ has a lower value when including IMC effect.

\subsubsection{charged pion $\pi^+$}
With our fitted $G(eB)$ in Fig.\ref{geb}, we solve the gap equation (\ref{gap}) and pole equation (\ref{pip}) to obtain $\pi^+$ meson mass at finite temperature and magnetic field with IMC effect. The results with and without IMC effect are plotted in red and blue, respectively.

\begin{figure}[hbt]
\centering
\includegraphics[width=7cm]{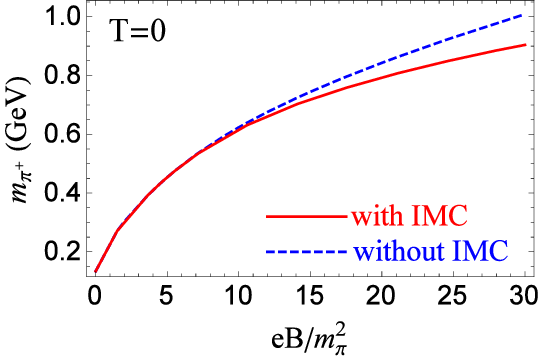}
\caption{$\pi^{+}$ meson mass $m_{\pi^+}$ as a function of magnetic field at vanishing temperature, with (red solid line) and without (blue dashed line) inverse magnetic catalysis effect.} \label{masspicharge}
\end{figure}

As shown in Fig.\ref{masspicharge}, with and without IMC effect, $m_{\pi^+}$ are increasing functions of magnetic field, and no decreasing behavior is observed at vanishing temperature. With IMC effect, $m_{\pi^+}$ has a lower value than without IMC effect, and the deviation becomes larger at stronger magnetic field. Similar results are obtained in three-flavor NJL model including IMC effect~\cite{su3meson4}.

\begin{figure}[hbt]
\centering
\includegraphics[width=7cm]{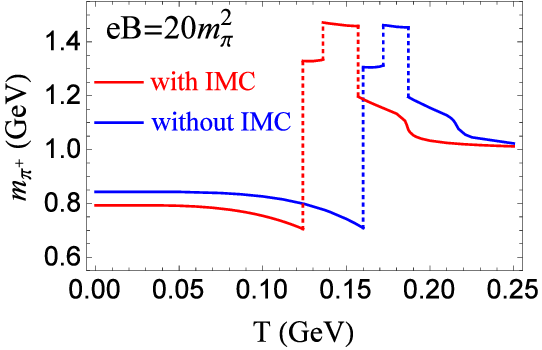}
\caption{$\pi^{+}$ meson mass $m_{\pi^+}$ as a function of temperature at fixed magnetic field $eB=20 m_{\pi}^2$, with (red lines) and without (blue lines) inverse magnetic catalysis effect.} \label{masspicharget}
\end{figure}

In Fig.\ref{masspicharget}, we make comparison of $m_{\pi^+}$ as a function of temperature at fixed magnetic field $eB=20 m_{\pi}^2$ with (red lines) and without (blue lines) inverse magnetic catalysis effect. They show similar structure. $m_{\pi^+}$ decreases in the low temperature region. At the Mott transition $T^+_m$, $m_{\pi^+}$ shows the first jump, and two other mass jumps happen at $T^+_1$ and $T^+_2$. With $T^+_m<T<T^+_1$, $m_{\pi^+}$ first decreases and then increases with temperature. With $T^+_1<T<T^+_2$ and $T>T^+_2$, $m_{\pi^+}$ decreases with temperature. At high enough temperature, $m_{\pi^+}$ becomes independent of temperature and IMC effect. In the low and high temperature regions, $m_{\pi^+}$ with IMC effect is smaller than without IMC effect. The value of $T^+_m=124, 160$ MeV, $T^+_1=136, 172$ MeV and $T^+_2=157, 187$ MeV are different for the cases with and without IMC effect, which indicates that IMC effect lowers down the temperatures of ${\pi^+}$ meson mass jumps.

\begin{figure}[hbt]
\centering
\includegraphics[width=7cm]{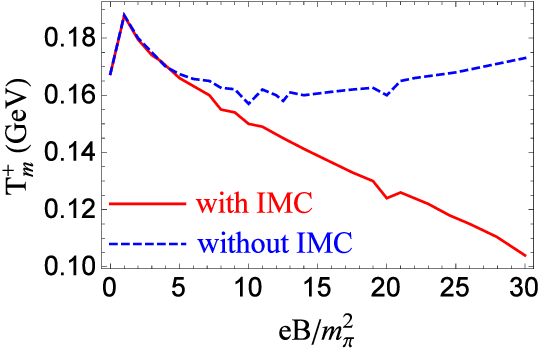}
\caption{Mott transition temperature of $\pi^{+}$ meson $T^+_m$ as a function of magnetic field, with (red solid line) and without (blue dashed line) inverse magnetic catalysis effect.} \label{tmottcharge}
\end{figure}

Figure \ref{tmottcharge} is the Mott transition temperature $T^+_m$ of $\pi^{+}$ meson as a function of magnetic field, with (red solid line) and without (blue dashed line) inverse magnetic catalysis effect. A fast increase of $T^+_m$ occurs when turning on external magnetic field, where a peak structure around $eB \simeq 1 m^2_{\pi}$ shows up. Without IMC effect, the $T^+_m$ decreases with magnetic field and then increases, which are associated with some oscillations. With IMC effect, the $T^+_m$ decreases as magnetic field goes up, which is also accompanied with some oscillations. At fixed magnetic field, $T^+_m$ has a lower value, when including IMC effect, .

\subsection{Current quark mass effect}
In this part, we consider the effect of current quark mass $m_0$ on mass spectra and Mott transition of charged pions. The results are plotted with black, blue, red, cyan and magenta lines, respectively, corresponding to $m_0=2,\ 5,\ 10,\ 20,\ 50$ MeV. In numerical calculations, we only change the parameter $m_0$ and keep other parameters intact in our NJL model.

\subsubsection{neutral pion $\pi^0$}
\begin{figure}[hbt]
\centering
\includegraphics[width=7cm]{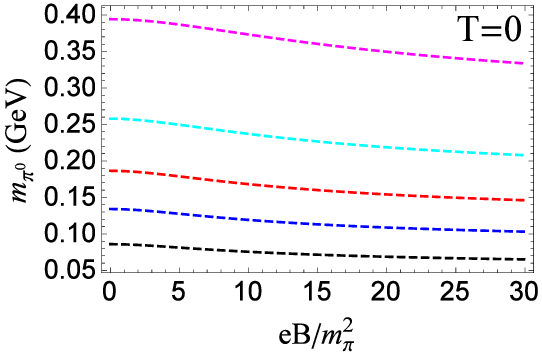}
\includegraphics[width=7cm]{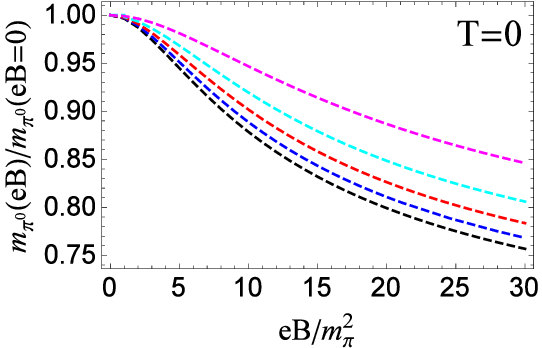}
\caption{$\pi^{0}$ meson mass $m_{\pi^0}$ (upper panel) and normalized $\pi^{0}$ meson mass $m_{\pi^0}(eB)/m_{\pi^0}(eB=0)$ (lower panel) as a function of magnetic field at vanishing temperature with different current quark mass $m_0$. The results are plotted with black, blue, red, cyan and magenta lines, respectively, corresponding to $m_0=2,\ 5,\ 10,\ 20,\ 50$ MeV.} \label{masspim0}
\end{figure}

Figure \ref{masspim0} plots the $\pi^{0}$ meson mass $m_{\pi^0}$ and normalized $\pi^{0}$ meson mass $m_{\pi^0}(eB)/m_{\pi^0}(eB=0)$ as a function of magnetic field at vanishing temperature with different current quark mass $m_0=2,\ 5,\ 10,\ 20,\ 50$ MeV. Because the current quark mass determines the explicit breaking of chiral symmetry. As the pseudo-Goldstone boson, the mass of $\pi^{0}$ meson will increase with current quark mass when fixing magnetic field and vanishing temperature. On the other side, magnetic field plays the role of catalysis for spontaneous breaking of chiral symmetry when fixing current quark mass and vanishing temperature, and this will lead to a decreasing $m_{\pi^0}$. As shown in Fig.\ref{masspim0}, with fixed magnetic field, $m_{\pi^0}$ becomes larger with larger $m_0$. Moreover, the deviation between $m_{\pi^0}$ with different $m_0$ looks independent of the magnetic field. With fixed current quark mass, $m_{\pi^0}$ is a decreasing function of magnetic field. Similar as $m_{\pi^0}$, the normalized $\pi^{0}$ meson mass $m_{\pi^0}(eB)/m_{\pi^0}(eB=0)$ increases with the current quark mass when fixing magnetic field, and decreases with magnetic field when fixing $m_0$, which are consistent with LQCD results~\cite{l4}.
\begin{figure}[hbt]
\centering
\includegraphics[width=7cm]{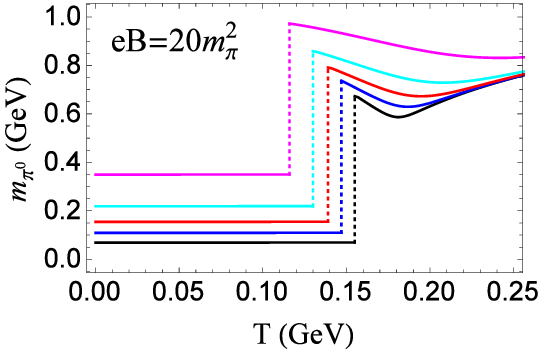}
\includegraphics[width=7cm]{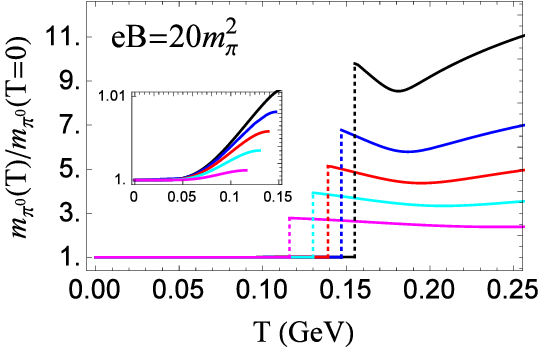}
\caption{$\pi^{0}$ meson mass $m_{\pi^0}$ (upper panel) and normalized $\pi^{0}$ meson mass $m_{\pi^0}(T)/m_{\pi^0}(T=0)$ (lower panel) as a function of temperature with fixed magnetic field $eB=20 m_{\pi}^2$ and different current quark mass $m_0$. The results are plotted with black, blue, red, cyan and magenta lines, respectively, corresponding to $m_0=2,\ 5,\ 10,\ 20,\ 50$ MeV.} \label{masspim0t}
\end{figure}

In Fig.\ref{masspim0t}, we depict the $\pi^{0}$ meson mass $m_{\pi^0}$ and normalized $\pi^{0}$ meson mass $m_{\pi^0}(T)/m_{\pi^0}(T=0)$ as a function of temperature with fixed magnetic field $eB=20 m_{\pi}^2$ and different current quark mass $m_0=2,\ 5,\ 10,\ 20,\ 50$ MeV. They show similar structure. $m_{\pi^0}$ slightly increases with temperature, and a mass jump happens at the Mott transition $T=T^{0}_m$. After Mott transition, $m_{\pi^0}$ first decreases and then increases with temperature. However, there exist some quantitative difference. With larger current quark mass, the $m_{\pi^0}$ is larger in the whole temperature region, and the Mott transition temperature is lower. At high enough temperature, $m_{\pi^0}$ will become degenerate due to the strong thermal motion of constituent quarks. Different from meson mass $m_{\pi^0}$, the normalized $\pi^{0}$ meson mass $m_{\pi^0}(T)/m_{\pi^0}(T=0)$ is larger with smaller current quark mass, and the deviation between different $m_0$ is larger with higher temperature.

\begin{figure}[hbt]
\centering
\includegraphics[width=7cm]{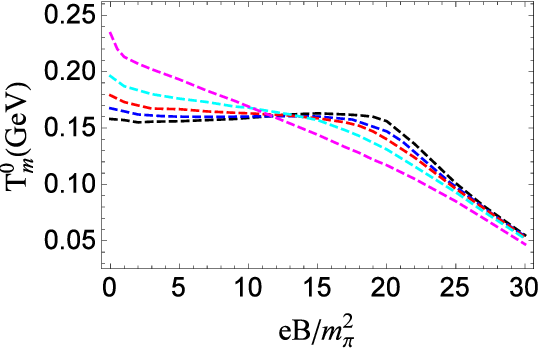}
\caption{Mott transition temperature of $\pi^{0}$ meson $T^{0}_m$ as a function of magnetic field with different current quark mass $m_0$. The results are plotted with black, blue, red, cyan and magenta lines, respectively, corresponding to $m_0=2,\ 5,\ 10,\ 20,\ 50$ MeV.} \label{tmottp0m0}
\end{figure}

The Mott transition temperature of $\pi^{0}$ meson $T^{0}_m$ is plotted as a function of magnetic field in Fig.\ref{tmottp0m0} with different current quark mass $m_0=2,\ 5,\ 10,\ 20,\ 50$ MeV. For a lower value of current quark mass $m_0=2$ MeV, the Mott transition temperature decreases with weak magnetic field $(eB\leq 2 m^2_{\pi})$, slightly increases in medium magnetic field region $(2 m^2_{\pi} \leq eB\leq 15 m^2_{\pi})$ and decreases again in strong magnetic field region $(eB\geq 15 m^2_{\pi})$. For $m_0=5$ MeV, we obtain a flat curve for the Mott transition temperature in medium magnetic field region $(5 m^2_{\pi} \leq eB \leq 16 m^2_{\pi})$, and in weak and strong magnetic field regions, the Mott transition temperature decreases. With the larger value of current quark mass $m_0=10,\ 20,\ 50$ MeV, the Mott transition temperatures are monotonic decreasing functions of magnetic field. In weak magnetic field region, $T^{0}_m$ is higher for larger $m_0$, but in strong magnetic field region, $T^{0}_m$ is lower for larger $m_0$.

In the end of this section, we make some comment on chiral limit with $m_0=0$. As discussed in our previous paper~\cite{mao1}, in chiral limit, the Goldstone boson $\pi^0$ is massless in chiral breaking phase and its mass continuously increases with temperature in chiral restored phase. The Mott transition temperature is the same as the critical temperature of chiral restoration phase transition, and it increases with magnetic field. Moreover, no mass jump occurs for $\pi^{0}$ meson at the Mott transition with or without external magnetic field.

\subsubsection{charged pion $\pi^+$}
\begin{figure}[hbt]
\centering
\includegraphics[width=7cm]{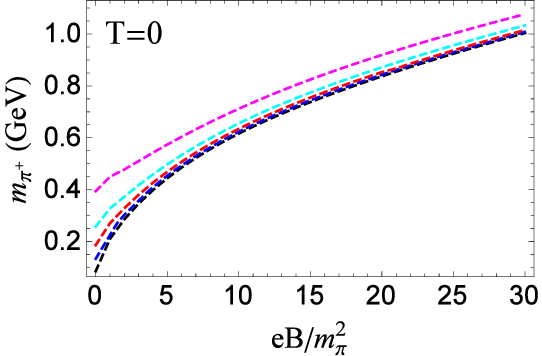}
\includegraphics[width=7cm]{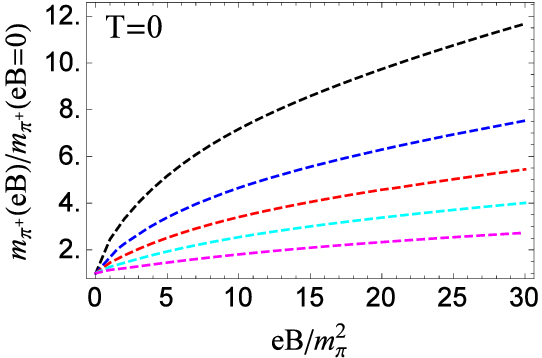}
\caption{$\pi^{+}$ meson mass $m_{\pi^+}$ (upper panel) and normalized $\pi^{+}$ meson mass $m_{\pi^+}(eB)/m_{\pi^+}(eB=0)$ (lower panel) as a function of magnetic field with vanishing temperature and different current quark mass $m_0$. The results are plotted with black, blue, red, cyan and magenta lines, respectively, corresponding to $m_0=2,\ 5,\ 10,\ 20,\ 50$ MeV.} \label{masschargepim0}
\end{figure}

Figure \ref{masschargepim0} shows $\pi^{+}$ meson mass $m_{\pi^+}$ and normalized mass $m_{\pi^+}(eB)/m_{\pi^+}(eB=0)$ as a function of magnetic field with vanishing temperature and different current quark mass $m_0=2,\ 5,\ 10,\ 20,\ 50$ MeV. $m_{\pi^+}$ is an increasing function of magnetic field when fixing $m_0$.  With fixed magnetic field, larger value of $m_0$ leads to larger $m_{\pi^+}$. In weak magnetic field region, the $m_0$ effect is more obvious than in strong magnetic field region, which is indicated by the larger difference of $m_{\pi^+}$ between different $m_0$ cases. However, with fixed magnetic field, the normalized $\pi^{+}$ meson mass $m_{\pi^+}(eB)/m_{\pi^+}(eB=0)$ decreases as $m_0$ goes up, which is consistent with LQCD results~\cite{l4}.

\begin{figure}[hbt]
\centering
\includegraphics[width=7cm]{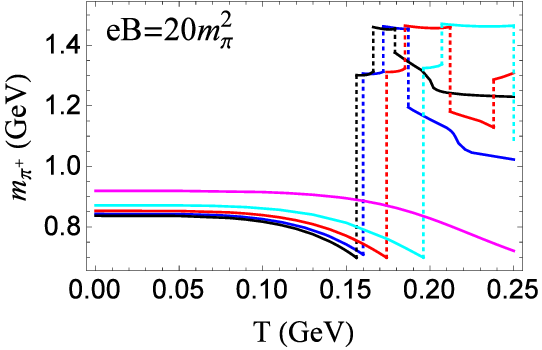}
\caption{$\pi^{+}$ meson mass $m_{\pi^+}$ as a function of temperature with fixed magnetic field $eB=20m^2_{\pi}$ and different current quark mass $m_0$. The results are plotted with black, blue, red, cyan and magenta lines, respectively, corresponding to $m_0=2,\ 5,\ 10,\ 20,\ 50$ MeV.} \label{fmpichargem0}
\end{figure}

$\pi^{+}$ meson mass $m_{\pi^+}$ is plotted as a function of temperature with fixed magnetic field $eB=20m^2_{\pi}$ and different current quark mass $m_0=2,\ 5,\ 10,\ 20,\ 50$ MeV in Fig.\ref{fmpichargem0}. $m_{\pi^+}$ has several mass jumps for all considered current quark mass $m_0$, which happen at temperature $T_m^+$, $T_1^+$, $T_2^+$, $T_3^+$ successively. With low temperature $T<T_m^+$, $m_{\pi^+}$ becomes larger with larger $m_0$. With $T_m^+<T<T_1^+$ and $T_1^+<T<T_2^+$, $m_{\pi^+}$ are also larger with larger $m_0$, but the value of $m_{\pi^+}$ is very close for different $m_0$. With $T>T_2^+$, $m_{\pi^+}$ changes nonmonotonically with $m_0$. It is clearly shown that the temperatures with mass jump depend on current quark mass. Within the considered temperature region $0<T<0.25$ GeV, we observe three jumps in case of current quark mass $m_0=2,\ 5,\ 20$ MeV, four jumps in case of $m_0=10$ MeV. For $m_0=50$ MeV, the first mass jump occurs at $T=0.262$ GeV, which is beyond the scope of Fig.\ref{fmpichargem0}.

\begin{figure}[hbt]
\centering
\includegraphics[width=7cm]{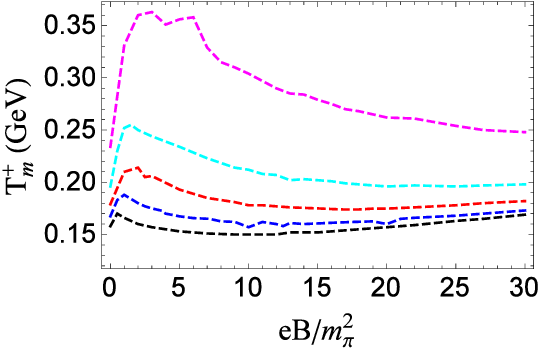}
\caption{Mott transition temperature of $\pi^{+}$ meson $T^+_m$ as a function of magnetic field with different current quark mass $m_0$. The results are plotted with black, blue, red, cyan and magenta lines, respectively, corresponding to $m_0=2,\ 5,\ 10,\ 20,\ 50$ MeV.} \label{tmottchargem0}
\end{figure}

Figure \ref{tmottchargem0} plots Mott transition temperature of $\pi^{+}$ meson $T^+_m$ as a function of magnetic field with different current quark mass $m_0=2,\ 5,\ 10,\ 20,\ 50$ MeV. $T^+_m$ shows similar behavior when varying current quark mass. Accompanied with some oscillations, $T^+_m$ increases fast in weak magnetic field region, decreases in medium magnetic field region and slightly increases in strong magnetic field region. With larger current quark mass, Mott transition temperature $T^+_m$ becomes higher. Note that in case of $m_0=50$ MeV, the increasing behavior of $T^+_m$ appears in very strong magnetic field, which is beyond the scope of Fig.\ref{tmottchargem0}.

\section{summary}
\label{sec:s}

Mass spectra and Mott transition of pions $(\pi^0,\ \pi^\pm)$ at finite temperature and magnetic field are investigated in a two-flavor NJL model, and we focus on the inverse magnetic catalysis effect and current quark mass effect.

We consider the inverse magnetic catalysis effect by introducing a magnetic dependent coupling constant into NJL model, which is a monotonic decreasing function of magnetic field. The mass spectra of pions $(\pi^0,\ \pi^\pm)$ at finite temperature and/or magnetic field are not changed qualitatively by IMC effect. At the Mott transition, the mass jumps of pions happen. Without IMC effect, the Mott transition temperature of $\pi^{0}$ meson $T^0_m$ decreases with magnetic field, but shows a flat structure in medium magnetic field region. With IMC effect, the flat structure of $T^0_m$ disappears and $T^0_m$ is a monotonic decreasing function of magnetic field. For charged pions $\pi^{\pm}$, the Mott transition temperature $T^+_m$ is not a monotonic function of magnetic field. Without IMC effect, it fast increases in weak magnetic field region, decreases in the medium magnetic field region and slightly increases in strong magnetic field region, which are accompanied with some oscillations. When including IMC effect, the increasing behavior of $T^+_m$ in strong magnetic field is changed into a decreasing behavior.

The current quark mass effect is studied in non-chiral limit. The masses of pions $(\pi^0,\ \pi^\pm)$ at vanishing temperature increases as the current quark mass $m_0$ goes up. However, the normalized masses $m_{\pi}(eB)/m_{\pi}(eB=0)$ change differently. For $\pi^{0}$ meson, $m_{\pi^0}(eB)/m_{\pi^0}(eB=0)$ increases with $m_0$. For $\pi^{+}$ meson, $m_{\pi^+}(eB)/m_{\pi^+}(eB=0)$ decreases with $m_0$. These properties are consistent with LQCD simulations. At the Mott transition, the mass jumps of pions happen. The Mott transition temperature of $\pi^{0}$ meson $T^0_m$ is qualitatively modifies by current quark mass effect. With a low value of current quark mass $m_0=2$ MeV, $T^0_m$ decreases with weak magnetic field, slightly increases in the medium magnetic field region, and decreases again in the strong magnetic field region. With $m_0=5$ MeV, we obtain a flat curve for $T^0_m$ in the medium magnetic field region, and in the weak and strong magnetic field region, $T^0_m$ decreases. With a larger value of current quark mass, such as $m_0=10,\ 20,\ 50$ MeV, $T^0_m$ is a monotonic decreasing functions of magnetic field. In the weak magnetic field region, the Mott transition temperature is higher for larger $m_0$, but in the strong magnetic field region, the Mott transition temperature is lower for larger $m_0$. For $\pi^{+}$ meson, the Mott transition temperature $T^+_m$ is only quantitatively modifies by current quark mass effect. Associated with some oscillations, $T^+_m$ increases fast in the weak magnetic field region, decreases in the medium magnetic field region and slightly increases in the strong magnetic field region. With larger $m_0$, $T^+_m$ becomes higher.

Due to the interaction with magnetic field, the charged pions $\pi^\pm$ show different behavior from the neutral pion $\pi^0$. One common character of pions $(\pi^0,\ \pi^\pm)$ is the mass jump at their Mott transitions, which is induced by the dimension reduction of the constituent quarks, and independent on the IMC effect and CQM effect. As a consequence of such jumps, it may result in some interesting phenomena in relativistic heavy ion collisions where the strong magnetic field can be created. For instance, when the formed fireball cools down, there might be sudden enhancement/suppression of pions.\\

\noindent {\bf Acknowledgement:}
Dr. Luyang Li is supported by Natural Science Basic Research Plan in Shaanxi Province of China (Program No.2023-JC-YB-570). Prof. Shijun Mao is supported by the NSFC Grant 12275204 and Fundamental Research Funds for the Central Universities.\\

\end{document}